\documentclass[prd,twocolumn,showpacs,preprintnumbers,amsmath,amssymb,superscriptaddress,floatfix,nofootinbib]{revtex4-2}
\usepackage{graphicx}
\usepackage{amsmath}
\usepackage{amsfonts}
\usepackage{amssymb}
\usepackage{color}
\usepackage{multirow}
\usepackage[colorlinks, citecolor=blue,anchorcolor=red,menucolor=red,linkcolor=red,filecolor=red,runcolor=red,urlcolor=blue,frenchlinks=red]{hyperref}

\usepackage{subfigure}

\begin{document}    

\title{Unveiling the $a_0(1710)$ nature in the process $J/\psi \to  {\bar{K}}^0K^+\rho^- $}

\date{\today}
\author{Yan Ding}
\affiliation{School of Physics, Zhengzhou
	University, Zhengzhou 450001, China}

\author{En Wang}\email{wangen@zzu.edu.cn}
\affiliation{School of Physics, Zhengzhou
	University, Zhengzhou 450001, China}
\affiliation{Guangxi Key Laboratory of Nuclear Physics and Nuclear Technology, Guangxi Normal University, Guilin 541004, China}

\author{De-Min Li}\email{lidm@zzu.edu.cn}
\affiliation{School of Physics, Zhengzhou
	University, Zhengzhou 450001, China}

\author{Li-Sheng Geng} \email{lisheng.geng@buaa.edu.cn}
\affiliation{School of Physics, Beihang University, Beijing 102206,
China}
\affiliation{Beijing Key Laboratory of Advanced Nuclear Materials and Physics, Beihang University, Beijing, 102206, China}
\affiliation{Peng Huanwu Collaborative Center for Research and Education, Beihang University, Beijing 100191, China}
\affiliation{Southern Center for Nuclear-Science Theory (SCNT), Institute of Modern Physics, Chinese Academy of Sciences, Huizhou 516000, Guangdong Province, China}

\author{Ju-Jun Xie} \email{xiejujun@impcas.ac.cn}
\affiliation{Southern Center for Nuclear-Science Theory (SCNT), Institute of Modern Physics, Chinese Academy of Sciences, Huizhou 516000, Guangdong Province, China}
\affiliation{Institute of Modern Physics, Chinese Academy of Sciences, Lanzhou 730000, China} 
\affiliation{School of Nuclear Sciences and Technology, University of Chinese Academy of Sciences, Beijing 101408, China}

\begin{abstract}
We have investigated the process $J/\psi \to {\bar{K}}^0K^+\rho^-$ by taking into account the $S$-wave ${K^*\bar{K}^*}$, $\rho \omega$, and $\rho \phi$ final-state interactions, where the scalar meson $a_0(1710)$ is generated. In addition, we also take into account the contributions from the scalar $a_0(980)(\to \bar{K}^0K^+)$ and the intermediate resonances $K_1(1270)^{-}(\to {\bar{K}}^0\rho^-) $ and $K_1(1270)^{0}(\to K^+\rho^-)$. Our results show that, in the ${{\bar{K}}^0K^+}$ invariant mass distribution, a clear peak structure around 1.8~GeV appears, which could be associated with the scalar  $a_0(1710)$, however, no significant structure of the $a_0(980)$ is observed.
On the other hand, one can find clear peaks of the $K_1(1270)$ in the ${\bar{K}}^0\rho^-$ and $K^+\rho^-$ invariant mass distributions. The future precise measurement of this process by the BESIII and Belle II Collaborations and  the planned Super Tau-Charm Facility (STCF) in the future could shed light on the nature of $a_0(1710)$.

\end{abstract}
\maketitle

\section{Introduction} \label{sec:Introduction}
In the last two decades, many states have been accumulated experimentally whose properties cannot be well described by the $q\bar{q}$ mesons and the $qqq$ baryons within the conventional quark model. Some exotic explanations are proposed for their nature, such as tetraquark, pentaquark, hybrid, glueball, kinematic effects, and the mixing of different components~\cite{Chen:2016spr,Chen:2022asf,Guo:2017jvc,Oset:2016lyh,Liu:2024uxn,Gao:1999ar,Wang:2024jyk}.  It is difficult to distinguish between those explanations, especially for states with the quantum numbers allowed by the conventional quark model~\cite{Guo:2017jvc,Wang:2024jyk}. 

Recently, the {\it BABAR} Collaboration reported the scalar resonance $a_0(1710)$ in the $\pi^\pm\eta$ invariant mass spectrum of the process $\eta_c\to \eta\pi^+\pi^-$~\cite{BaBar:2021fkz}. The $a_0(1710)$ state was also observed by the BESIII Collaboration in the $K_S^0K_S^0$ invariant mass spectrum of the process $D_s^+ \to K_S^0 K_S^0\pi^+$~\cite{BESIII:2021anf} and in the  $K_S^0K^+$ invariant mass spectrum of the process $D_s^+ \to K_S^0 K^+\pi^0$~\cite{BESIII:2022npc}.
We have tabulated the experimental masses and widths of $a_0(1710)$ in Table~\ref{tab:exp.mass and width}. It should be noted that, in Ref.~\cite{BESIII:2021anf}, BESIII did not distinguish between the $a_0(1710)$ and $f_0(1710)$ in the process $D_s^+ \to K_S^0 K_S^0\pi^+$, and denoted the combined state as $S(1710)$, while in Ref.~\cite{BESIII:2022npc} the $a_0(1710)$ was renamed as $a_0(1817)$ because of the different fitted Breit-Wigner mass of this state.

\begin{table}[htbp]
	\begin{center}
		\caption{\label{tab:exp.mass and width} Experimental measurements on the mass ($M_{a_0(1710)}$) and width ($\Gamma_{a_0(1710)}$) of the scalar state $a_0(1710)$. The first error is statistical, and the second one is systematic. All values are in units of MeV.}		
		\begin{tabular}{cccc}\hline\hline
			
			Collaboration & $M_{a_0(1710)}$ \quad & $\Gamma_{a_0(1710)}$ \quad &Ref.  \\ \hline
			$\textit{BABAR}$  \quad  & $1704 \pm 5 \pm 2$ \quad  &\quad $110\pm 15\pm 11$ &~\cite{BaBar:2021fkz}  \quad       \\
			BESIII      \quad     & $1723 \pm 11 \pm 2$ \quad  &\quad $140 \pm 14 \pm 4$ &~\cite{BESIII:2021anf} \quad     \\
			BESIII     \quad    &  $1817\pm 8 \pm 20$ \quad  &\quad
			$97 \pm 22 \pm 15$ &~\cite{BESIII:2022npc} \quad    \\
			\hline\hline			
		\end{tabular}
	\end{center}
\end{table}

Before the observation of $a_0(1710)$, there have been many theoretical studies about the $a_0(1710)$ and its isospin partner $f_0(1710)$ from various perspectives~\cite{Nagahiro:2008bn,Branz:2009cv,Geng:2010kma,Xie:2014gla,MartinezTorres:2012du,Wang:2021jub,Garcia-Recio:2010enl,Garcia-Recio:2013uva,Close:2005vf,Gui:2012gx,Janowski:2014ppa,Fariborz:2015dou,Chen:2005mg,Achasov:2023zoi,Achasov:2023izs}. In Refs.~\cite{Geng:2008gx,Du:2018gyn}, the $f_0(1710)$, as a well-established state according to the Review of Particle Physics (RPP)~\cite{ParticleDataGroup:2022pth}, could be dynamically generated from the vector-vector interactions, and one isovector scalar state $a_0$ with a mass around 1770~MeV was also predicted, the picture of which remains essentially the same when the pseudoscalar-pseudoscalar coupled channels were taken into account~\cite{Wang:2022pin}. 
Based on the SU(6) spin-flavor symmetry, the $f_0(1710)$ is mostly a $K^*\bar{K}^*$ bound state, and an $a_0$ state with a pole position of $\sqrt{s_R}=(1760,-12)$~MeV was predicted to couple strongly to $K^*\bar{K}^*$ and $\phi\rho$ in Ref.~\cite{Garcia-Recio:2010enl}. 
In addition, a scalar state $a_0$ with a mass of 1744~MeV is also predicted in the approach of the Regge trajectories~\cite{Wang:2017pxm}.
In Ref.~\cite{Close:2005vf}, it was suggested that the $f_0(1710)$ wave function contains a large $s\bar{s}$ component, while in Refs.~\cite{Gui:2012gx,Janowski:2014ppa,Fariborz:2015dou,Chen:2005mg}, the $f_0(1710)$ was regarded as the candidate of  a scalar glueball. Recently, it was shown that the $a_0(1710)$ as a $K^*\bar{K}^*$ molecule plays an important role in the three-body interactions of $\eta K^*\bar{K}^*$, which could dynamically generate the $\pi(2070)$~\cite{Shen:2023uus}.

As shown in Table~\ref{tab:exp.mass and width},  there is not yet a consensus on the mass of the $a_0(1710)$ experimentally, which could complicate understanding its nature. For instance, $a_0(1710)$ [or $a_0(1817)$] and $X(1812)$ have been explained as the $3^{3}P_0$  $q\bar{q}$ state by assuming $a_0(980)$ and $f_0(980)$ as $1^{3}P_0$ $q\bar{q}$ states~\cite{Guo:2022xqu}. However, $X(1812)$ was observed in the process $J/\psi \to \gamma\phi\omega$ by the BESIII Collaboration~\cite{BES:2006vdb,BESIII:2012rtd}, and the enhancement near the $\phi\omega$ threshold, associated with $X(1812)$, could be described by the reflection of $f_0(1710)$, as discussed in Ref.~\cite{MartinezTorres:2012du}. Regarding the $a_0(1710)$ as a $K^*\bar{K}^*$ molecular state, Refs.~\cite{Zhu:2022wzk,Zhu:2022guw,Dai:2021owu,Oset:2023hyt,Wang:2023aza}  have successfully described  the invariant mass distributions of the processes $D_s^+ \to K_S^0 K_S^0\pi^+$ and  $D_s^+ \to K_S^0 K^+\pi^0$ measured by the BESIII Collaboration~\cite{BESIII:2021anf,BESIII:2022npc}. 

Since the peak positions of the $a_0(1710)$ in the $K\bar{K}$ invariant mass distributions of the processes $D_s^+\to K_S^0 K_S^0 \pi^+, K_S^0K^+\pi^0$ observed by the BESIII Collaboration are very close to the boundary region of the $K\bar{K}$ invariant mass, we have suggested to measure its properties in the process $\eta_c \to \bar{K}^0K^+\pi^-$ in Ref.~\cite{Ding:2023eps}, and  predicted a dip structure around 1.8~GeV, associated with the $a_0(1710)$, in the $\bar{K}^0K^+$ invariant mass distribution~\cite{Ding:2023eps}, which is consistent with the {\it BABAR} measurements~\cite{BaBar:2015kii}. In addition, the photoproduction process is also proposed to search for the $a_0(1710)$ in Ref.~\cite{Wang:2023lia}.

The BESIII Collaboration has accumulated $(10.09 \pm 0.04) \times 10^9$ $J/\psi$ events at the BEPCII
collider~\cite{BESIII:2021cxx}, and the Super Tau-Charm Facility (STCF) project under development in China is expected to accumulate $3.4\times 10^{12}$ $J/\psi$ events per year~\cite{Achasov:2023gey}. Since the dominant decay mode of the $a_0(1710)$ state is $K\bar{K}$ within the molecular picture~\cite{Geng:2008gx,Wang:2022pin}, 
it is natural to search for the scalar $a_0(1710)$ in the process $J/\psi\to a_0(1710)^+\rho^-\to \bar{K}^0K^+\rho^-$.
It should be stressed that the {\it BABAR} Collaboration has measured this process, and the branching fraction is $\mathcal{B}(J/\psi\to K_S^0K^{\pm}\rho^{\mp})=(1.87\pm0.18\pm 0.34 ) \times 10^{-3}$~\cite{BaBar:2017pkz}. However, the  $K_S^0K^{\pm}$ mass spectrum was not reported by the {\it BABAR} Collaboration~\cite{BaBar:2017pkz}.

In this work, we will investigate the process $J/\psi \to a_0(1710)^+/a_0(980)^+\rho^- \to \bar{K}^0K^+\rho^-$ by considering the contributions from the intermediate resonances $a_0(1710)$ and $a_0(980)$.
It should be pointed out that another state $a_2(1700)$ with a mass of $1706\pm 14$~MeV and a width of $378^{+60}_{-50}$~MeV may give a broad contribution and should not significantly affect the narrow peak of $a_0(1710)$ in the $\bar{K}^0K^+$ invariant mass distribution~\cite{ParticleDataGroup:2022pth}. Furthermore,  $a_2(1700)$ couples to $K\bar{K}$ in the $D$ wave with a small branching fraction of $\mathcal{B}(a_2(1700)\to K\bar{K})=(1.3\pm 0.8)\%$. Therefore, we will neglect its contribution.

 On the other hand, the interactions of vector mesons and pseudoscalar mesons within the unitary chiral approach could dynamically generate the resonance $K_1(1270)$ with a two-pole structure~\cite{Geng:2006yb,Wang:2020pyy,Wang:2019mph}, where the lower pole mainly couples to the $K^*\pi$ channel, and the higher one couples strongly to the $K\rho$ channel. Thus, we also consider the contribution from the intermediate resonance $K_1(1270)$ in this work.
Considering that the branching ratios of the process $K_1(1400)\to K\rho$ and $K^*(1410)\to K\rho$ are $(3.0\pm 3.0)\%$ and $<7\%$, which are 10 times smaller than the $\mathcal{B}(K_1(1270)\to K\rho)=(38\pm 13)\%$~\cite{ParticleDataGroup:2022pth}, we will also neglect the contributions from the intermediate states $K_1(1400)$ and $K^*(1410)$ in this work. 
Thus, the precise measurements of the process $J/\psi\to \bar{K}^0K^+\rho^-$ could also shed light on the two-pole structure of $K_1(1270)$, which is crucial to understanding the hadron-hadron interactions~\cite{Xie:2023jve,Xie:2023cej}.

The paper is organized as follows. In Sec.~\ref{sec:Formalism}, we present the
theoretical formalism for studying the $J/\psi \to \bar{K}^0 K^+\rho^-$ decay, and
in Sec.~\ref{sec:Results}, we show our numerical results and discussions,
followed by a summary in the last section.

\section{Formalism} \label{sec:Formalism}

First, we present the theoretical formalism for studying the process $J/\psi \to \bar{K}^0 K^+\rho^-$ via the $K^*\bar{K}^*$, $\omega\rho$, and $\phi\rho$ final-state interactions in coupled channels, which will generate the scalar resonance $a_0(1710)$ in Sec.~\ref{subsec:2A}. Next, we show the formalism for the process  $J/\psi \to K_1(1270)^- K^{+} [K_1(1270)^0 \bar{K}^{0}]$  with $K_1(1270)^- \to \bar{K}^{0} \rho^-$ [$K_1(1270)^{0} \to K^+ \rho^-$] in Sec.~\ref{Subsec:2B}. In the Sec.~\ref{Subsec:2C}, we will describe the contribution from the $S$-wave $K\bar{K}$ final-state interaction, which would generate the scalar meson $a_0(980)$. At last, the formalism of the double differential widths for this process is given in Sec.~\ref{Subsec:2D}.

\subsection{Mechanism for the intermediate $a_0(1710)$}\label{subsec:2A}
\begin{figure}[tbhp]
    \centering
        \includegraphics[scale=0.38]{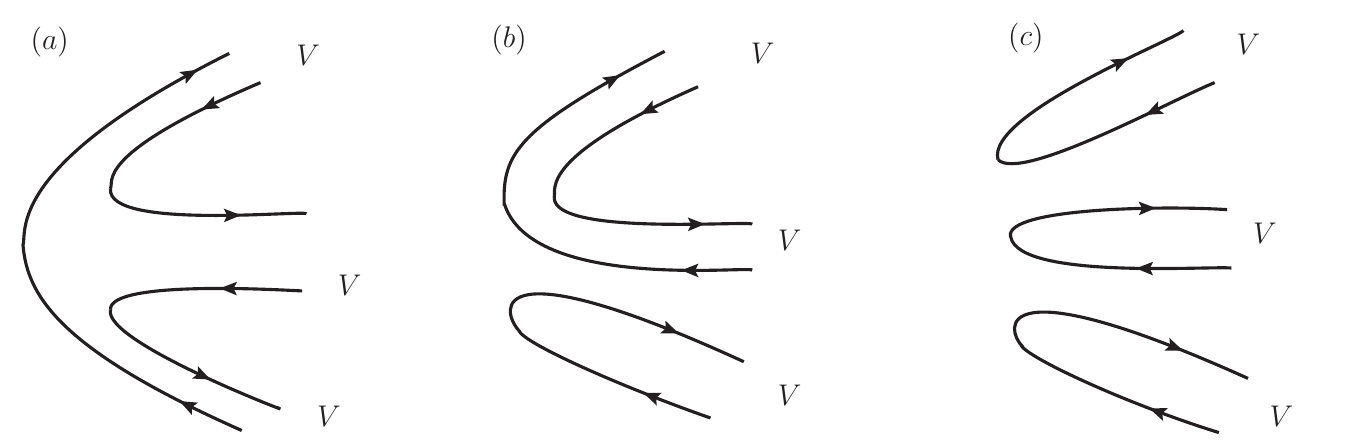}
    \caption{Diagrammatic expression of (a) $\textless VVV\textgreater$, (b) $\quad \textless VV\textgreater \textless V\textgreater$, (c) $\quad\textless V\textgreater\textless V\textgreater\textless V\textgreater$ terms in Eq.~\eqref{eq:2}.}
  \label{fig:traceV}
\end{figure}
As done in Refs.~\cite{Zhu:2022guw,Zhu:2022wzk,Ding:2023eps}, the scalar meson $a_0(1710)$ is regarded as a vector-vector molecular state~\cite{Geng:2008gx,Du:2018gyn}. To study the role of $a_0(1710)$ in the $J/\psi \to \rho^- K^+ \bar{K}^0$ decay, one needs to first produce the meson $\rho^-$ and a vector-vector pair, then the final-state interactions of the vector-vector pair will produce the $a_0(1710)$, which decays into $K^+ \bar{K}^0$ in the final state. Considering that the $J/\psi$ is a flavor singlet, we could introduce the combination modes in the primary vertex~\cite{Ikeno:2019grj,Jiang:2019ijx,Sakai:2019uig}, whose diagrammatic expression is depicted in Fig.~\ref{fig:traceV},
\begin{eqnarray}
\textless VVV\textgreater,\quad \textless VV\textgreater \textless V\textgreater, \quad\textless V\textgreater\textless V\textgreater\textless V\textgreater,\label{eq:2}
\end{eqnarray}
where $V$ is the matrix of the SU(3) vector mesons~\cite{Ikeno:2019grj,Jiang:2019ijx,Duan:2021pll},
\begin{eqnarray}
V =\left(\begin{matrix} \frac{{\rho}^0}{\sqrt{2}} + \frac{{\omega}}{\sqrt{2}}  & \rho^+  & K^{*+}  \\
		\rho^-  &   - \frac{{\rho}^0}{\sqrt{2}} + \frac{{\omega}}{\sqrt{2}}  &  K^{*0} \\
		K^{*-}  &  \bar{K}^{*0}   &   \phi
	\end{matrix}
	\right),
\end{eqnarray}
where the symbol \textless...\textgreater stands for the  trace of the SU(3) matrices. Since no term contains $\rho^-$ in $\textless V\textgreater\textless V\textgreater\textless V\textgreater$, we do not take this combination in our work. One could obtain the relevant contributions by isolating the terms containing $\rho^-$, as follows:
\begin{eqnarray}
&&\textless VVV\textgreater:\alpha \times \left[\frac{\rho^+\rho^0}{\sqrt{2}}+3\sqrt{2}\omega\rho^+ + 3\bar{K}^{*0}K^{\ast+}\right]\rho^-, \\ 
&&\textless VV\textgreater \textless V\textgreater: \beta \times \left[2\sqrt{2}\omega\rho^++2\phi\rho^+\right]\rho^-.\label{eq:3}
\end{eqnarray} 
Here, two parameters  $\alpha$ and $\beta$ are introduced to account for the weights of $\textless VVV\textgreater$ and $\textless VV\textgreater \textless V\textgreater$ structures, respectively. It is concluded in Refs.~\cite{Abreu:2023xvw,Geng:2008gx,Molina:2019wjj} that the $\textless VVV\textgreater$ one was favored and the best ratio $\beta/\alpha=0.32$ was obtained by fitting to the experimental measurements of the process $J/\psi \to \phi V V$~\cite{Geng:2008gx,Molina:2019wjj}, thus we will use this finding and take $\alpha=1$ and  $\beta=0.32$ in this work.


\begin{figure}[tbhp]
    \centering
        \includegraphics[scale=0.8]{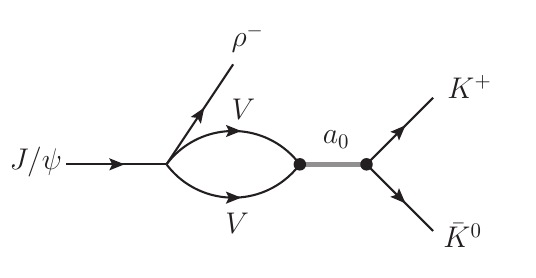}
    \caption{Diagram for the process $J/\psi \to VV \rho^- \to a_0(1710)^+ \rho^- \to K^+ \bar{K}^0 \rho^-$, where $VV$ stands for $\bar{K}^{*0}K^{\ast+}$, $\omega \rho^+$, and $\phi\rho^+$.}
  \label{fig:hadronFSI}
\end{figure}

In the molecular picture, the $a_0(1710)$ is dynamically generated from the $S$-wave $K^*\bar{K}^*$, $\omega\rho$, and $\phi\rho$ interactions in coupled channels~\cite{Geng:2008gx,Du:2018gyn}~\footnote{It should be noted that, the conservation of $G$ parity forbids the coupling of $a_0(1710)$ to the channel $\rho\rho$.}, and then decays into the final state $\bar{K}^0K^+$, as depicted in Fig.~\ref{fig:hadronFSI}.
The decay amplitude of Fig.~\ref{fig:hadronFSI} can be written as,
{\begin{eqnarray}
\mathcal{M}_a &=& V_p \times \left [ 3\alpha G_{\bar{K}^{*0}K^{\ast+}}t_{\bar{K}^{*0}K^{\ast+} \to \bar{K}^0K^+} \right. \nonumber\\
&& \left.+ (2\sqrt{2}\beta+3\sqrt{2}\alpha ) G_{\omega\rho^+}t_{\omega\rho^+ \to \bar{K}^0K^+}
\right. \nonumber\\
&& \left.+2\beta G_{\phi\rho^+}t_{\phi\rho^+ \to \bar{K}^0K^+}
\right ], \label{eq:ma}
\end{eqnarray}}
where $V_p$ is a global factor, and $t_{\bar{K}^{*0}K^{\ast+} \to \bar{K}^0K^+}$, $t_{\omega\rho^+ \to \bar{K}^0K^+}$, and $t_{\phi\rho^+ \to \bar{K}^0K^+}$ are the transition amplitudes. ${G}_{\bar{K}^{*0}K^{\ast+}}$, ${G}_{\omega\rho^+}$, and $G_{\phi\rho^+}$ are the loop functions for the $\bar{K}^{*0}{K}^{*+}$, $\omega\rho^+$, and $\phi\rho^+$ channels, respectively, and read~\cite{Geng:2008gx,Molina:2008jw},
\begin{eqnarray}
	{G}_{i}(M_{\bar{K}^0K^+}) &=& \int_{m_{1-}^2}^{m_{1+}^2} \int_{m_{2-}^2}^{m_{2+}^2}d\tilde{m}_1^2d\tilde{m}_2^2 \times \nonumber \\  
	&& \!\!\!\!\!\!\!\!\!\!\!\!\!\!\!\!\!\!\!\! \omega(\tilde{m}_1^2)\omega(\tilde{m}_2^2)\tilde{G}(M_{\bar{K}^0K^+},\tilde{m}_1^2,\tilde{m}_2^2), \label{eq:loop_vector}
\end{eqnarray}
where
\begin{gather}
\omega(\tilde{m}_i^2) = {\frac{1}{N}}\left(-\frac{1}{\pi}\right)\text{Im}\left[\frac{1}{\tilde{m}_i^2-m_{V_{i}}^2+i\Gamma(\tilde{m}_i^2)\tilde{m}_i}\right],\\
N = \int_{\tilde{m}_{i-}^2}^{\tilde{m}_{i+}^2}d\tilde{m}_i^2\left(-\frac{1}{\pi}\right)\text{Im}\left[\frac{1}{\tilde{m}_i^2-m_{V_{i}}^2+i\Gamma(\tilde{m}_i^2)\tilde{m}_i}\right], \\
\Gamma(\tilde{m}_i^2) = \Gamma_{V_{i}}\frac{\tilde{k}^3}{k^3}\Theta(\tilde{m}-m_{P_1}-m_{P_2}), \\
	\tilde{k} = \frac{\lambda^\frac{1}{2}(\tilde{m}_i^2,m_{P_{1}}^2,m_{P_{2}}^2)}{2\tilde{m}_i},~	k = \frac{\lambda^\frac{1}{2}(m_{V_i}^2,m_{P_{1}}^2,m_{P_{2}}^2)}{2m_{V_i}},
\end{gather}
with the K${\ddot{a}}$llen function $\lambda(x,y,z) =x^2+y^2+z^2-2xy-2xz-2yz$.
Here, we consider the decay channels $\pi\pi$ and $K\pi$ for the vector mesons $\rho$ and $K^*$, respectively, and neglect the small widths of $\omega$ ($\Gamma_\omega=8.68$~MeV) and $\phi$ ($\Gamma_\phi=4.249$~MeV). Taking the vector $K^{\ast}$ for example, $m_{1+}^2=\left(m_{K^\ast}+2\Gamma_{K^\ast}\right)^2$ and $m_{1-}^2=\left(m_{K^\ast}-2\Gamma_{K^\ast}\right)^2$. Similarly, one can obtain $m_{1+}^2$ and $m_{1-}^2$ for the $\rho$ meson. The masses, widths, and spin parities of the involved particles are taken from the RPP~\cite{{ParticleDataGroup:2022pth}}, as  listed in Table~\ref{tab:particleparameters}.

\begin{table}[htbp]
\caption{Masses, widths, and spin-parities of the involved particles in this work. All values are in units of MeV.}	\label{tab:particleparameters}
	\begin{tabular}{cccc}
		\hline\hline  
		State & Mass & Width  & Spin parity ($J^P$) \\ \hline
		$J/\psi$  &3096.900    &0.0926               &$1^-$      \\
		$\rho^{\pm,0}$  &775.26   &149.1             &$1^-$   \\
		$\bar{K}^0$  &497.611     &-            &$0^-$ \\
        $K^\pm$      &493.677      &-           &$0^-$ \\
         $K^{*}$  &893.6     &49.1          &$1^-$ \\
        $\omega$  &782.66       &8.68           &$1^-$ \\
        $\phi$   &1019.461      &4.249           &$1^-$ \\
        $K_1(1270)$ &1284    &146           &$1^+$ \\
        		\hline\hline
	\end{tabular}
\end{table}

The loop function $\tilde{G}$ in Eq.~(\ref{eq:loop_vector}) is for stable particles, and in the dimensional regularization scheme, it can be written as~\cite{Lyu:2023ppb},
\begin{eqnarray}
	\tilde{G} &=& \frac{1}{16\pi^2}\Bigg\{a_{\mu} + \text{ln}\frac{m_1^2}{\mu^2}+\frac{m_2^2-m_1^2+s}{2s}\text{ln}\frac{m_2^2}{m_1^2} \nonumber \\
	&& \frac{p}{\sqrt{s}}\bigg[\text{ln}\left(s-\left(m_2^2-m_1^2\right)+2p\sqrt{s}\right)\nonumber \\
	&& +\text{ln}\left(s+\left(m_2^2-m_1^2\right)+2p\sqrt{s}\right)\nonumber \\
	&& -\text{ln}\left(-s+\left(m_2^2-m_1^2\right)+2p\sqrt{s}\right)\nonumber \\
	&& -\text{ln}\left(-s-\left(m_2^2-m_1^2\right)+2p\sqrt{s}\right)\bigg]\Bigg\},\label{eq:loop-stable}
\end{eqnarray}
with
\begin{eqnarray}
	p=\frac{\lambda^{1/2}(s,m_1^2,m_2^2)}{2\sqrt{s}},
\end{eqnarray}
where $a_{\mu}$ is the subtraction constant, $\mu$ is the dimensional regularization scale, and $s=M^2_{\bar{K}^0K^+}$. In this work, we take $a_{\mu} =-1.726$  and $\mu = 1000$~MeV as used in Ref.~\cite{Geng:2008gx}. It is worth mentioning that any change in $\mu$ could be reabsorbed by a change in $a_\mu$ through $a_{\mu'} - a_\mu = {\rm ln}(\mu'^{2}/\mu^2)$, which implies that the loop function $\tilde{G}$ is scale independent~\cite{Duan:2022upr}. 

\begin{figure}[htbp]
	\centering
	\includegraphics[scale=0.7]{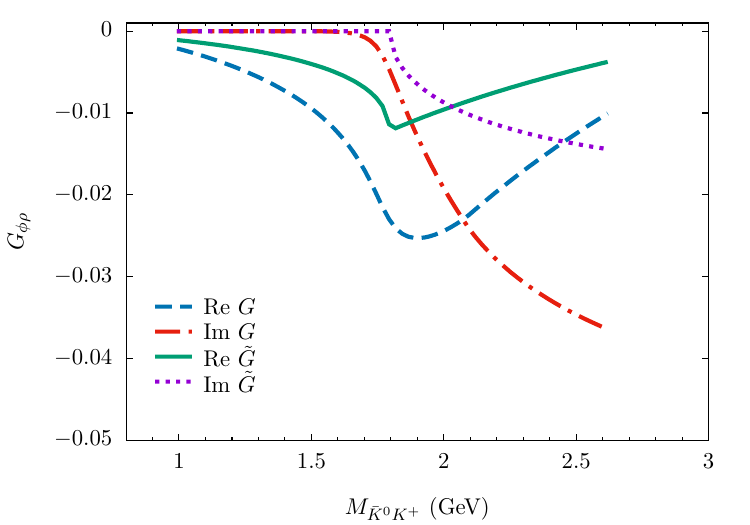}
	\vspace{-0.2cm}
	\caption{Real and imaginary parts of the loop functions $G_{\phi\rho}$ and $\tilde{G}_{\phi\rho}$ as a function of the $\bar{K}^0K^+$ invariant mass. The blue dashed and red dot-dashed curves correspond to the real and imaginary parts of the loop function $G$, considering the width of $\rho$. Meanwhile, the green solid and purple dotted curves correspond to the real and imaginary parts of the loop function $\tilde{G}$ without the contribution from the $\rho$ width, respectively.}
	\label{fig:Gphirho}
\end{figure}

To show the influence of the widths of vector mesons on the loop functions,  we have calculated the loop functions $G_{\phi\rho^+}$ and $\tilde{G}_{\phi\rho^+}$ as functions of the $\bar{K}^0K^+$ invariant mass,\footnote{One can find the invariant mass distributions of the loop functions $\tilde{G}_{\bar{K}^{*0}K^{\ast+}}$ and $\tilde{G}_{\omega\rho}$ in Refs.~\cite{Zhu:2022wzk,Ding:2023eps}.}  as presented in Fig.~\ref{fig:Gphirho}. The blue dashed and red dot-dashed curves correspond to the real and imaginary parts of the loop function $G$, considering the width of $\rho$. In contrast, the green solid and purple dotted curves correspond to the real and imaginary parts of the loop function $\tilde{G}$ without the contribution from the $\rho$ width, respectively. One can find that the loop functions $G$, considering the width of the vector meson, become smoother around the threshold of the $\phi\rho$.

\begin{table}[htbp]
	\caption{Mass, width, and coupling constants of the scalar $a_0(1710)$~\cite{Geng:2008gx}. All values are in units of MeV.}	\label{tab:parameters}
	\begin{tabular}{cc}
		\hline\hline  
		 Parameters & Value \\ \hline
      Mass &   1777 \\
      Width &  148 \\
      $\Gamma_{K\bar{K}}$  & 36 \\
      $g_{K\bar{K}}$ &1966 \\
      $g_{\rho\rho}$ &0 \\
       $g_{K^*{\bar{K}}^*}$ & (7525, -1529)  \\
        $g_{\omega\rho}$ &  (-4042, 1391)  \\
   $g_{\phi\rho}$&  (4998, -1872)    \\ \hline
   \vspace{0.05cm}
	\end{tabular}
\end{table}

On the other hand, the transition amplitudes $t_{i \to \bar{K}^0K^+}$ in Eq.~\eqref{eq:ma} can be written as,
\begin{eqnarray}
t_{i \to \bar{K}^0K^+}={\frac{g_{i} \times   g_{K\bar{K}}}{M_{\bar{K}^0K^+}^2-M_{a_0(1710)}^2+iM_{a_0(1710)}\Gamma_{a_0(1710)}}},
\end{eqnarray}
where $M_{a_0(1710)}$ and $\Gamma_{a_0(1710)}$ are the mass and width of the $a_0(1710)$, respectively, and we take their values from Refs.~\cite{Geng:2008gx,Geng:2009gb}, which are tabulated in Table~\ref{tab:parameters}. $g_i$ are the coupling constants of $a_0(1710)$ to $K^*{\bar{K}}^*$, $\omega\rho$, and $\phi\rho^+$, whose values are determined in Ref.~\cite{Geng:2008gx}, while the coupling $g_{K\bar{K}}$ is determined from the partial decay width of $a_0(1710) \to K\bar{K}$,
\begin{eqnarray}
\Gamma_{K\bar{K}} = \frac{g^2_{K\bar{K}}}{8\pi} \frac{|{\vec{p}}_K|}{M^2_{a_0}},
\end{eqnarray}
where ${\vec{p}}_K$ is the three-momentum of the $K$ or $\bar{K}$ meson in the $a_0(1710)$ rest frame,
\begin{align}
	|{\vec{p}}_K|=\frac{\lambda^{1/2}(M_{a_0}^2,m_{\bar{K}}^2,m_K^2)}{2M_{a_0}}.
\end{align}
With the partial decay width $\Gamma_{K\bar{K}}=36$~MeV predicted by Ref.~\cite{Geng:2008gx}, one can only obtain the absolute value of the coupling constant, but not the phase, thus by assuming that $g_{K\bar{K}}$ is real and positive we take $g_{K\bar{K}}=1966$~MeV, as done in Refs.~\cite{Zhu:2022wzk,Zhu:2022guw}.

\subsection{Mechanism for the intermediate $K_1(1270)$}
\label{Subsec:2B}

\begin{figure}[htbp]
	\centering
	\includegraphics[scale=0.6]{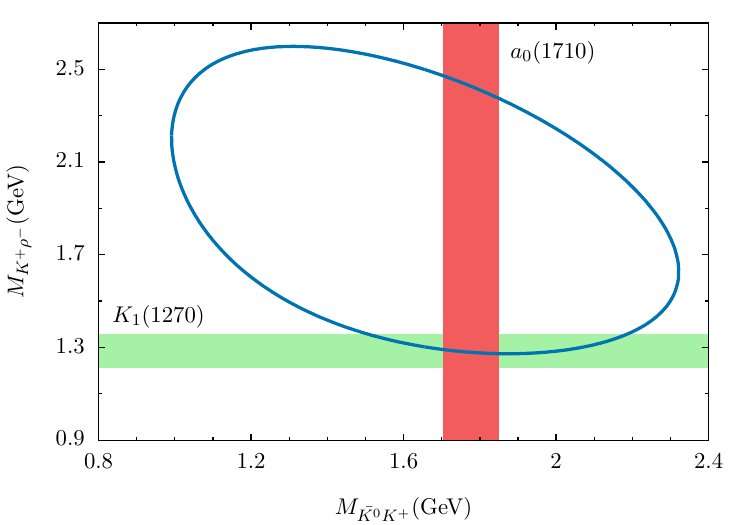}
	\vspace{-0.1cm}
	\caption{The Dalitz plot for the $J/\psi \rightarrow  \bar{K}^0K^+\rho^-$. The red band stands for the region of $M_{a_0}\pm \frac{1}{2}\Gamma_{a_0}$ where the predicted $a_0(1710)$ state lies. The green band stands for the region of $M_{K_1}\pm \frac{1}{2}\Gamma_{K_1}$ where the $K_1(1270)$ state lies.}
	\label{fig:2d}
\end{figure}

\begin{figure}[htbp]
	\centering
	\includegraphics[scale=0.9]{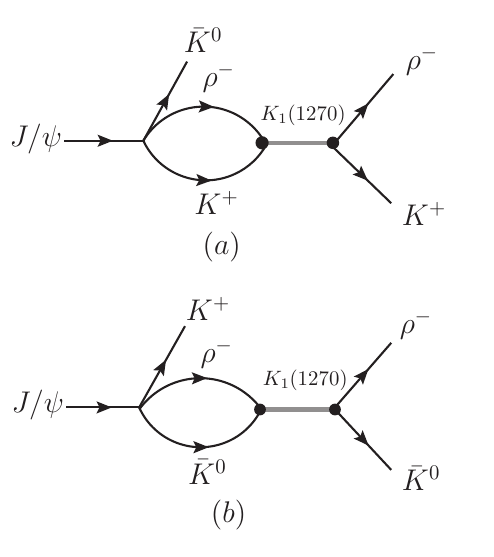}
	\caption{Diagrams for $J/\psi \rightarrow  \bar{K}^0K^+\rho^-$ via the intermediate (a) $K_1(1270)^0$  and (b) $K_1(1270)^-$, followed by the decay $K_1(1270)^{0,-} \to K^+\rho^- / \bar{K}^0\rho^-$ .}	\label{fig:K1}
\end{figure}

In Fig.~\ref{fig:2d}, we show the Dalitz plot for the process $J/\psi \rightarrow \bar{K}^0K^+\rho^-$, 
and one can find that the contribution from the $K_1(1270)$ (the green band) could interfere with the one from the $a_0(1710)$ (the red band) close to the  $K\rho$ threshold.
Since the $K_1(1270)$ state could be dynamically generated from the interaction of vector mesons and pseudoscalar mesons~\cite{Geng:2006yb,Wang:2020pyy,Wang:2020pyy}, the $K^+\rho^-$ and $\bar{K}^0\rho^-$ could undergo the $S$-wave final-state interaction, which will generate the $K_1(1270)$ state, followed by the decay $K_1(1270)\to K\rho$, as depicted in Fig.~\ref{fig:K1}.

\begin{table}[htbp]
	\caption{Pole positions and coupling constants of the two poles of the $K_1(1270)$~\cite{Geng:2006yb}. All values are in units of MeV.}	\label{tab:K1}
\vspace{0.07cm}
	\begin{tabular}{ccc}
		\hline\hline 
		 & First pole  \quad & Second pole  \\  \hline
              Pole position $\sqrt{s_0}$  & $1195-i123$ \quad & $1284-i73$ \\ 
   $g_{K\rho}$ & $-1671+i1599$  \quad & $4804+i395$\\ \hline
   \vspace{0.05cm}
	\end{tabular}
\end{table}

 The decay amplitude for $J/\psi\rightarrow \bar{K}^0 K_1(1270)^-\to \bar{K}^0K^+\rho^-$ of Fig.~\ref{fig:K1}(a) can be written as 
\begin{eqnarray}
    \mathcal{M}_b&=&V_p^{\prime} \times G_{K^+\rho^-}t_{K^+\rho^- \to K^+\rho^-} \label{eq:mb},
\end{eqnarray}
where $V'_p$ stands for the weight of the direct production vertex, and the $t_{K^+\rho^- \to K^+\rho^-}$ is the transition amplitude, which can be written as  
\begin{eqnarray}
t_{K^+\rho^- \to K^+\rho^-}&=&{\frac{g_{K^+\rho^-} g_{K^+\rho^-}}{M_{K^+\rho^-}^2-M_{K_1}^2+iM_{K_1}\Gamma_{K_1}}} \label{eq:mb},
\end{eqnarray}
where $M_{K^+\rho^-}$ is the invariant mass of the $K^+\rho^-$ system, and $g_{K^+\rho^-}$ denotes the coupling constant. In this work, we adopt the value $g_{K^+\rho^-}=(4804+i395)$~MeV of Table~\ref{tab:K1}. Since the higher pole of the $K_1(1270)$ mainly couples to the $K\rho$ channel, we can relate the mass and width of the $K_1(1270)$ with the higher pole position of Table~\ref{tab:K1}, i.e. $M_{K_1}={\rm Re}\sqrt{s_0}$ and $\Gamma_{K_1}=2{\rm Im}\sqrt{s_0}$.
Similarly, the amplitude of the process $J/\psi \to K^+ {K}_1(1270)^- \to K^+\bar{K}^0\rho^-$, as depicted in Fig.~\ref{fig:K1}(b), can be expressed as,
\begin{eqnarray}
	\mathcal{M}_c&=&V_p^{\prime} \times G_{\bar{K}^0\rho^-}t_{\bar{K}^0\rho^- \to \bar{K}^0\rho^-} \label{eq:mb},
\end{eqnarray}
\begin{eqnarray}
	t_{\bar{K}^0\rho^- \to \bar{K}^0\rho^-}&=&{\frac{g_{\bar{K}^0\rho^-} g_{\bar{K}^0\rho^-}}{M_{\bar{K}^0\rho^-}^2-M_{K_1}^2+iM_{K_1}\Gamma_{K_1}}}
 \label{eq:mc},
\end{eqnarray}
where $M_{\bar{K}^0\rho^-}$ is the $\bar{K}^0\rho^-$ invariant mass, and $g_{\bar{K}^0\rho^-}$ is the coupling constant, $g_{\bar{K}^0\rho^-}=g_{K^+\rho^-}=(4804+i395)$~MeV in this work. We take the same weight $V'_p$ for the contributions from the $K_1(1270)^0$ and $K_1(1270)^-$.

\begin{figure}[htbp]
	\centering
	\includegraphics[scale=0.6]{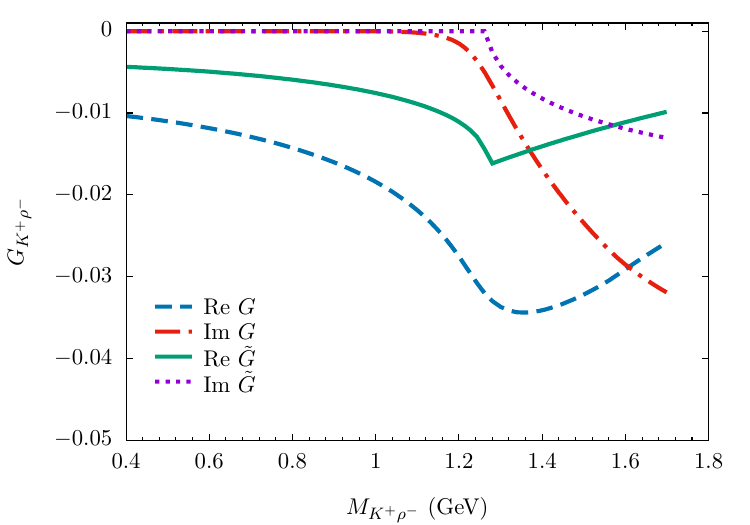}
	\vspace{-0.1cm}
	\caption{Real and imaginary parts of the loop functions $G_{K^+\rho^-}$ and $\tilde{G}_{K^+\rho^-}$ as a function of the $K^+\rho^-$ invariant  mass. The notations of the curves are the same as those of Fig.~\ref{fig:Gphirho}.}
	\label{fig:GKzrho}
\end{figure}

\begin{figure}[htbp]
	\centering
	\includegraphics[scale=0.6]{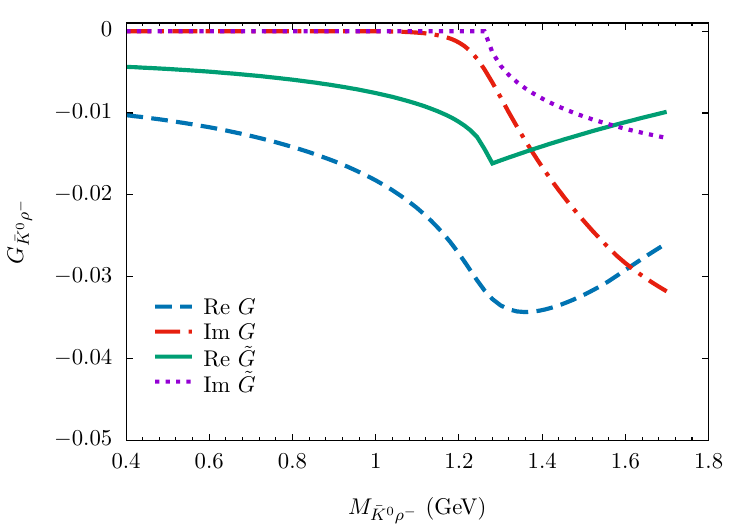}
	\vspace{-0.1cm}
	\caption{Real and imaginary parts of the loop functions $G_{\bar{K}^0\rho^-}$ and $\tilde{G}_{\bar{K}^0\rho^-}$ as a function of the $\bar{K}^0\rho^-$ invariant mass. The notations of the curves are the same as those of Fig.~\ref{fig:Gphirho}.}
	\label{fig:GK0rho}
\end{figure}

According to Eqs.~(\ref{eq:loop_vector}) and (\ref{eq:loop-stable}), we have also calculated the loop function $G_{K^+\rho^-}$ / $G_{\bar{K}^0\rho^-}$ and $\tilde{G}_{K^+\rho^-}$ / $\tilde{G}_{\bar{K}^0\rho^-}$   as functions of the $K^+\rho^-$ and $\bar{K}^0\rho^-$ invariant masses, respectively, as presented in Figs.~\ref{fig:GKzrho} and \ref{fig:GK0rho}. One can find that the loop functions $G$ become smoother around the threshold when considering the width of the $\rho$. 

\subsection{Mechanism for the intermediate $a_0(980)$}\label{Subsec:2C}

\begin{figure}[tbhp]
    \centering
        \includegraphics[scale=0.75]{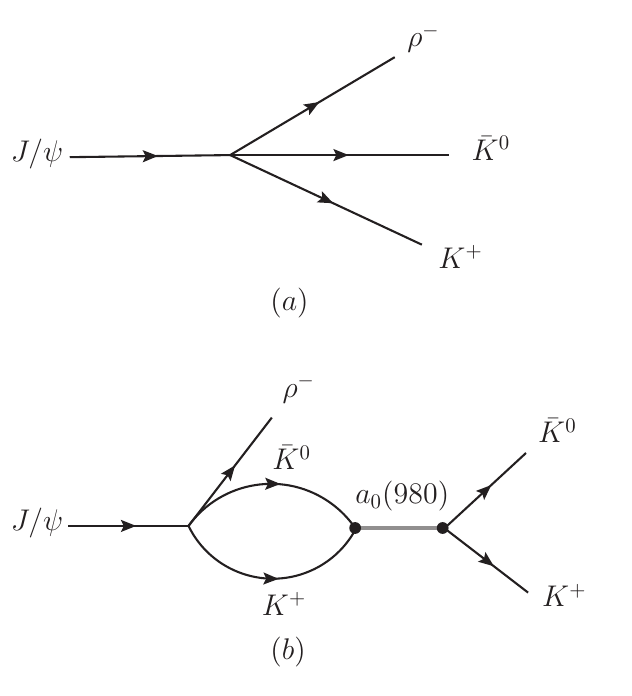}
    \caption{Diagrams for the process $J/\psi \to a_0(980)^+ \rho^- \to K^+ \bar{K}^0 \rho^-$. (a) Tree diagram and (b) the final-state interaction of $\bar{K}^0K^+$ to produce the $a_0(980)$ state.}
  \label{fig:980}
\end{figure}

In addition to the intermediate resonances $a_0(1710)$ and $K_1(1270)$, the process $J/\psi\to K^+ \bar{K}^0 \rho^-$ can happen via the direct production, as depicted in Fig.~\ref{fig:980}(a), and the $K^+\Bar{K}^0$ final-state interaction, which generates the scalar meson $a_0(980)$, as depicted in Fig.~\ref{fig:980}(b). Thus, the decay amplitude can be written as
\begin{eqnarray}
    \mathcal{M}_d&=&V_p^{\prime} \left[1+ G_{K\bar{K}} t_{\bar{K}^0K^+ \to \bar{K}^0K^+}\right] \label{eq:md},
\end{eqnarray}
where $V_p^{\prime}$ is the weight of the direct production vertex $J/\psi \to \rho^-\bar{K}^0K^+$ of Fig.~\ref{fig:980}(a), the same as the one of Eq.~(\ref{eq:mb}). The loop function $G_{K\bar{K}}$ could be  written as
\begin{eqnarray} 	
&&G_{K\bar{K}}=\\ \nonumber
&&i\int\frac{d^4q}{(2\pi)^4}\frac{1}{(P-q)^2-m_1^2+i\epsilon}\frac{1}{q^2-m_2^2+i\epsilon}, \label{eq:g-cutoff}
\end{eqnarray}
where $m_1$ and $m_2$ are the masses of the two mesons in the loop of the $K\bar{K}$ channel, and $P$ and $q$ are the four-momenta of the  $K\bar{K}$ system and the $\bar{K}$ meson, respectively. The loop function of Eq.~(\ref{eq:g-cutoff}) is logarithmically divergent. For $G_{K\bar{K}}$, we adopt the cutoff method, and perform the integral for $q$ in Eq.~(\ref{eq:g-cutoff}) with a cutoff $q_\text{max}$= 903~MeV to  regularize the
loop, the same as Ref.~\cite{Zhu:2022guw}.

The $t_{\bar{K}^0K^+}$ is the transition amplitude, which depends on the invariant mass $M_{\bar{K}^0K^+}$, and could be obtained in the chiral unitary approach by solving the Bethe-Salpeter equation~\cite{Oller:1997pn,Oller:1998hw,Oller:1997ng,Feng:2020jvp}
\begin{eqnarray}\label{eq:BS} 
T=[1-VG]^{-1}V 
\end{eqnarray}
where $V$ is a $2\times 2$ matrix with the transition potentials between the isospin channels $K\bar{K}$ and $\pi\eta$. The transition amplitudes $t_{\bar{K}^0K^+ \to \bar{K}^0K^+}$ in particle basis can be related to the one in isospin basis,
\begin{eqnarray} 
t_{\bar{K}^0K^+ \to \bar{K}^0K^+}=t_{K\bar{K} \to K\bar{K}}.
\end{eqnarray}

With the isospin multiplets $K=(K^+,K^0)$, $\bar{K}=(\bar{K}^0,-K^-)$, $\pi=(-\pi^+,\pi^0,\pi^-)$, the $2 \times 2$ matrix $V$ can be easily obtained as follows~\cite{Wang:2021naf,Xie:2014tma,Ling:2021qzl,Duan:2020vye}:
\begin{eqnarray} 
V_{K\bar{K} \to K\bar{K}}&=&-\frac{1}{4f^2}s \\ \nonumber
V_{K\bar{K} \to \pi\eta}&=&\frac{\sqrt{6}}{12f^2}(3s-\frac{8}{3}m_K^2-\frac{1}{3}m_{\pi}^2-m_{\eta}^2) \\ \nonumber
V_{\pi\eta \to K\bar{K}}&=&V_{K\bar{K} \to \pi\eta} \\ \nonumber
V_{\pi\eta \to \pi\eta}&=&-\frac{1}{3f^2}m_{\pi}^2,
\end{eqnarray}
where $f$ = 93~MeV is the pion decay constant, $m_{\pi}$ and $m_{K}$ are the
isospin averaged masses of the pion and kaon, respectively, and $s$ is the
invariant mass squared of the pseudoscalar-pseudoscalar
system.

It is worth mentioning that, since the transition amplitude of Eq.~(\ref{eq:BS}) is unstable in the high energy region, we should not use the model for higher invariant masses. For that purpose we take the following prescription: we evaluate $Gt(M_\text{inv})$ combinations up to $M_\text{inv}=M_\text{cut}$. From there on, we multiply $Gt$ by a smooth factor to make it gradually decrease at large $M_\text{inv}$. Thus, we take~\cite{Debastiani:2016ayp}
\begin{eqnarray} 
Gt(M_\text{inv})=Gt(M_\text{cut})e^{-\alpha(M_\text{inv}-M_\text{cut})}, \label{eq:decrease}
\end{eqnarray}
for 
\begin{eqnarray} 
M_\text{inv}>M_\text{cut}.
\end{eqnarray}
In our work, we take the value $M_\text{cut}$ = 1100~MeV and $\alpha=0.0037~\text{MeV}^{-1}$, as used in Ref.~\cite{Debastiani:2016ayp}.

\subsection{Invariant mass distributions}\label{Subsec:2D}

With the amplitudes obtained above, we can write down the total decay amplitude of $J/\psi \to \bar{K}^0K^+\rho^-$ as follows, 
\begin{eqnarray}\label{eq:fullamp}
 \mathcal{M} =\mathcal{M}_a+\mathcal{M}_b+\mathcal{M}_c+\mathcal{M}_d,
\end{eqnarray}
and the double differential widths of the  process $J/\psi \to \bar{K}^0K^+\rho^-$ are
\begin{eqnarray}
     \frac{d^2\Gamma}{dM_{\bar{K}^0K^+}{dM_{K^+\rho^-}}}=\frac{M_{\bar{K}^0K^+}M_{K^+\rho^-}}{128\pi^3m_{J/\psi}^3}|\mathcal{M}|^2, \label {eq:dwidth_nowidthofrho1} \\
     \frac{d^2\Gamma}{dM_{\bar{K}^0K^+}{dM_{\bar{K}^0\rho^-}}}=\frac{M_{\bar{K}^0K^+}M_{\bar{K}^0\rho^-}}{128\pi^3m_{J/\psi}^3}|\mathcal{M}|^2.\label {eq:dwidth_nowidthofrho2}
\end{eqnarray}
Furthermore, one can easily obtain ${d\Gamma}/{dM_{\bar{K}^0K^+}}$,  ${d\Gamma}/{dM_{\bar{K}^0 {\rho}^-}}$, and ${d\Gamma}/{dM_{K^+{\rho}^-}}$ by integrating over each of the invariant mass variables with the limits of the Dalitz plot given in the RPP~\cite{ParticleDataGroup:2022pth}. However, since the final meson $\rho^-$ has a large width ($\sim$149.1~MeV), and the $K_1(1270)$ mass is very close to the $K\rho$ threshold (as shown by Fig.~\ref{fig:2d}), one needs to take into account its finite width by folding with the vector-meson  $\rho^-$ spectral function for the invariant mass distributions~\cite{Wang:2019mph}, as follows,
\begin{eqnarray}
\frac{d\tilde{\Gamma}}{{d{M}_{12}}}=\int_{m_{\rho^-}-2\Gamma_{\rho^-}}^{m_{\rho^-}+2\Gamma_{\rho^-}}d\hat{m}_{\rho^-}\left[\frac{d\Gamma}{d{M}_{12}} \times \omega(\hat{m}^2_{\rho^-}) \right],\label {eq:dwidth_widthofrho}
\end{eqnarray}
where the mass $m_{\rho^-}$ in ${d\Gamma}/dM_{12}$ should be replaced by $\hat{m}_{\rho^-}$.
For example, the differential width ${d\tilde{\Gamma}}/{dM_{\bar{K}^0K^+}}$ of the  process $J/\psi \to \bar{K}^0K^+\rho^-$ becomes,
\begin{eqnarray}
     \frac{d\tilde{\Gamma}}{dM_{\bar{K}^0K^+}}&=& \int_{m_{\rho^-}-2\Gamma_{\rho^-}}^{{\rm min}\left(m_{\rho^-}+2\Gamma_{\rho^-},m_{J/\psi-M_{\bar{K}^0K^+}}\right)}d\hat{m}_{\rho^-}  \nonumber \\
     && \times\int_{{M_{K^+\rho^-}^{\rm min}}}^{M_{K^+\rho^-}^{\rm max}} dM_{K^+\rho^-}\nonumber  \\ &&\times\frac{M_{\bar{K}^0K^+}M_{K^+\rho^-}}{128\pi^3m_{J/\psi}^3}|\mathcal{M}|^2\times \omega(\hat{m}^2_{\rho^-}), \label {eq:dgammadm12dm23} 
\end{eqnarray}
where the range of $M_{\bar{K}^0K^+}$ is,
\begin{eqnarray}
    m_{\bar{K}^0}+m_{K^+}<M_{\bar{K}^0K^+}< m_{J/\psi}-m_{\rho^-}+2\Gamma_{\rho^-},
\end{eqnarray}
and the upper and lower limits for $M_{K^+\rho^-}$ are,
\begin{eqnarray}
\left(M_{K^+\rho^-}^{\rm max}\right)^2 &= &\left(E_{K^+}^\ast+E_{\rho^-}^\ast\right)^2 -  \nonumber \\
    && \left(\sqrt{E_{K^+}^{\ast2}-m_{K^+}^2}-\sqrt{E_{{\rho^-}}^{\ast2}-\hat{m}_{{\rho^-}}^2}\right)^2 \nonumber \\
\left(M_{K^+\rho^-}^{\rm min}\right)^2 &=&\left(E_{K^+}^\ast+E_{{\rho^-}}^\ast\right)^2 -  \nonumber \\    &&\left(\sqrt{E_{K^+}^{\ast2}-m_{K^+}^2}+\sqrt{E_{\rho^-}^{\ast2}-\hat{m}^2_{\rho^-}}\right)^2, \nonumber
\end{eqnarray}
where $E_{K^+}^\ast$ and $E_{\rho^-}^{\ast}$ are the energies of $K^+$ and $\rho^-$ in the $\bar{K}^0\rho^-$ rest frame, respectively,
\begin{align}
    &E_{K^+}^\ast=\frac{M_{\bar{K}^0K^+}^2-m_{\bar{K}^0}^2+m_{K^+}^2}{2M_{\bar{K}^0K^+}},  \nonumber \\
    &E_{\rho^-}^\ast=\frac{m_{J/\psi}^2-M_{\bar{K}^0K^+}^2-\hat{m}_{\rho^-}^2}{2M_{\bar{K}^0K^+}}.
\end{align}

\section{Results and Discussion} \label{sec:Results}

In our formalism, there are two unknown parameters: $V_p$ for the weight of the $a_0(1710)$ contribution and $V_p^{\prime}$ for the one of the intermediate $K_1(1270)$  and $a_0(980)$ contributions. Since there are some similarities between the processes $J/\psi \to VVV$ and $J/\psi \to VPP$, it is expected that $V_p$ and $V'_p$ are of the same order of magnitude. Thus, we first take $V_p=V'_p$ and discuss the influence on our results of different values of  $V_p$ and $V'_p$.

\begin{figure}[htbp]
\centering	\includegraphics[scale=0.65]{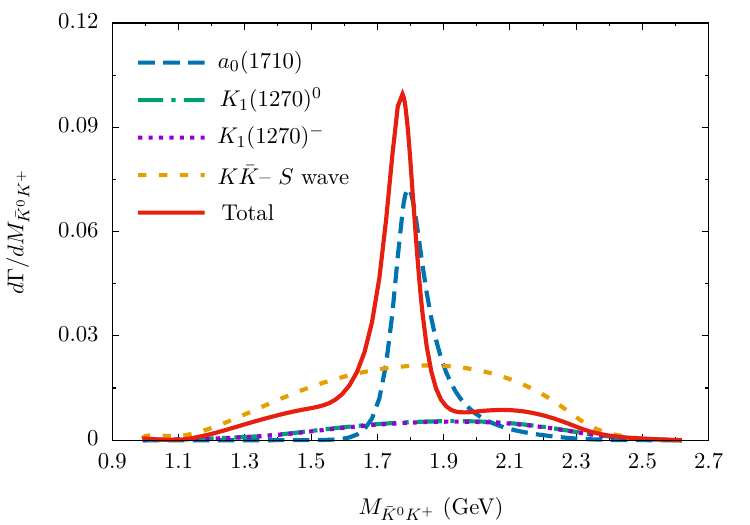} \caption{$\bar{K}^0K^+$ invariant mass distribution of the process $J/\psi\rightarrow \bar{K}^0K^+\rho^-$. The results are obtained with Eq.~(\ref{eq:dwidth_widthofrho}), where the width of final $\rho^-$ is considered. 
 The red-solid curve stands for the total contribution, while the blue-dashed curve, the green-dot-dashed curve, purple-dotted curve, and orange dashed curve correspond to the contributions from the $a_0(1710)$ state, the intermediate $K_1(1270)^0$, $K_1(1270)^-$, and the one from the  direction production and the $S$-wave $K\bar K$ interaction of Eq.~(\ref{eq:md}), respectively.}	\label{fig:kk}
\end{figure}

\begin{figure}[htbp]
\centering	\includegraphics[scale=0.65]{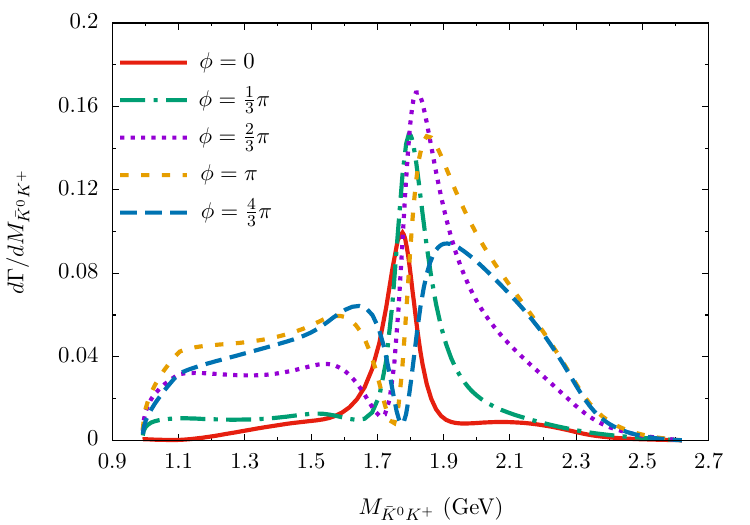} \caption{$\bar{K}^0K^+$ invariant mass distribution of the process $J/\psi\rightarrow \bar{K}^0K^+\rho^-$ obtained with a phase angle $\phi=0, \pi/3, 2\pi/3, \pi$, and $4\pi/3$, respectively. See the text for details.}	\label{fig:phi}
\end{figure}

\begin{figure}[htbp]
	\centering
\includegraphics[scale=0.65]{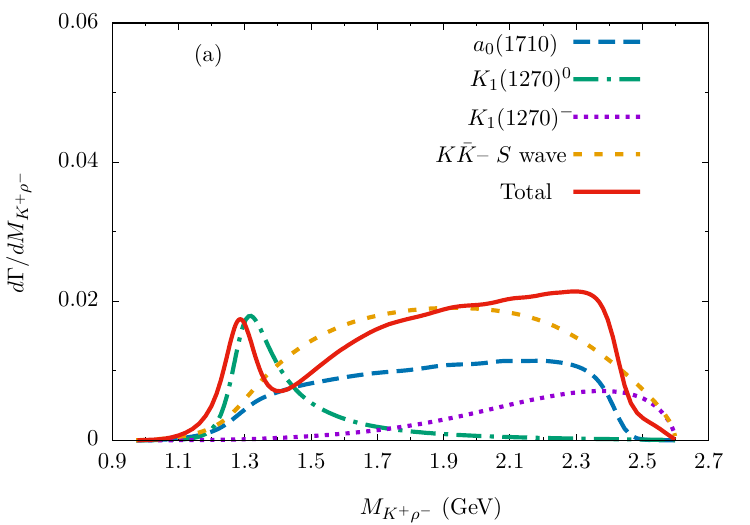}	
\includegraphics[scale=0.65]{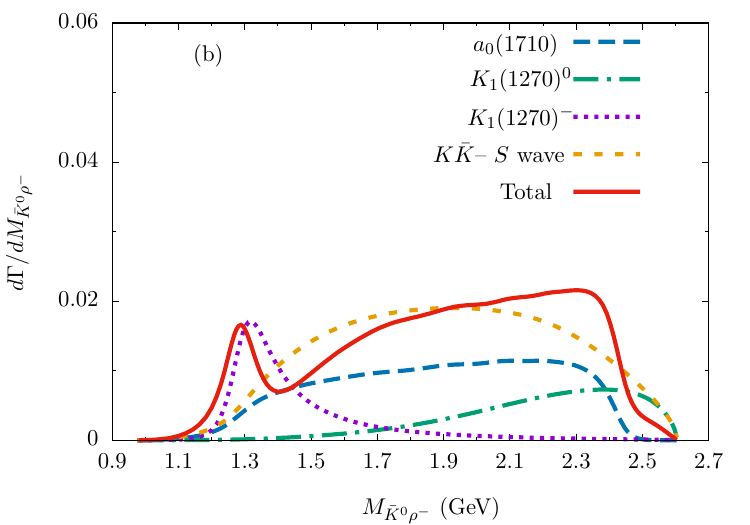}
	\caption{(a) $K^+\rho^-$ and (b) $\bar{K}^0\rho^-$ invariant mass distributions of the process $J/\psi\rightarrow \bar{K}^0K^+\rho^-$. The explanations of the curves are the same as those of Fig.~\ref{fig:kk}.}	\label{fig:dw_Krho}
\end{figure}

\begin{figure}[htbp]
	\centering
 \includegraphics[scale=0.65]{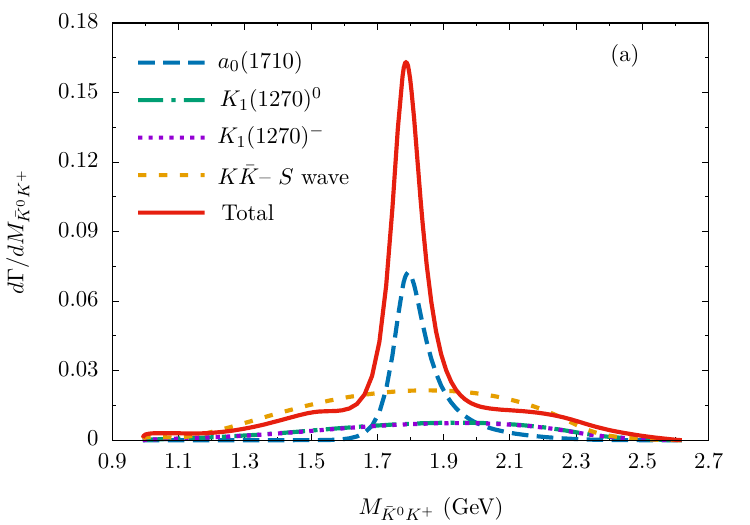} 
\includegraphics[scale=0.65]{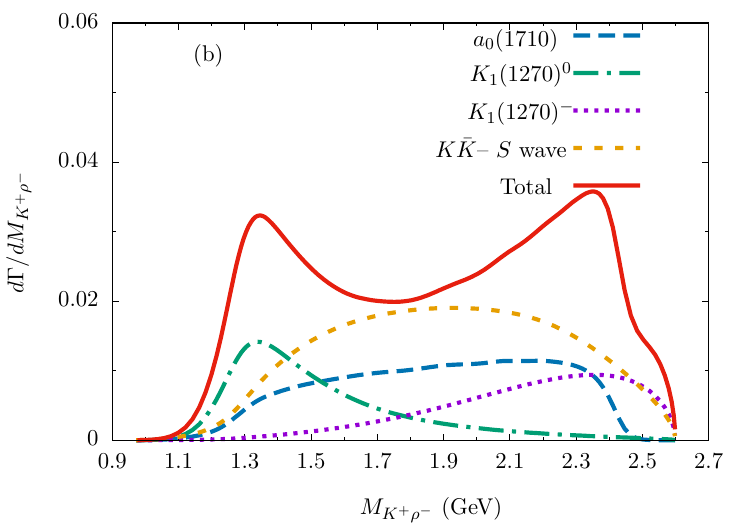}	
\includegraphics[scale=0.65]{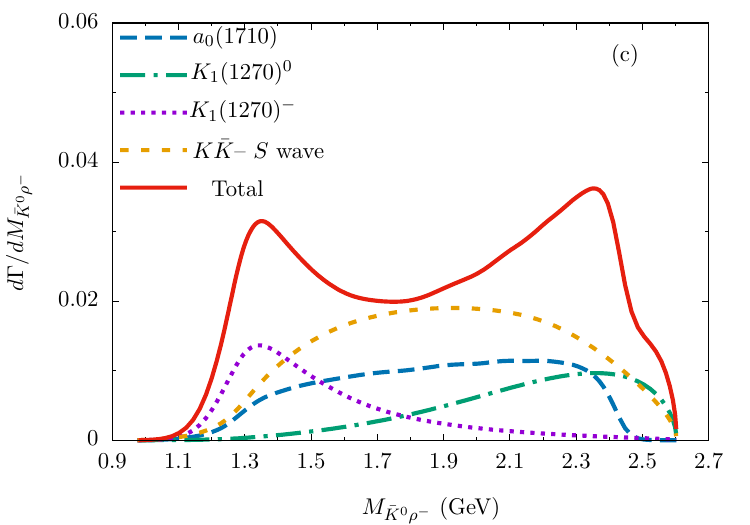}
	\caption{The (a) $\bar{K}^0K^+$, (b) $K^+\rho^-$ and (c) $\bar{K}^0\rho^-$  invariant mass distributions of the process $J/\psi\rightarrow \bar{K}^0K^+\rho^-$ with the parameters of lower pole of $K_1(1270)$. The explanations of the curves are the same as those of Fig.~\ref{fig:kk}.}	\label{fig:lowerpole-Krho}
 \end{figure}

\begin{figure}[htbp]
	\centering
	\includegraphics[scale=0.65]{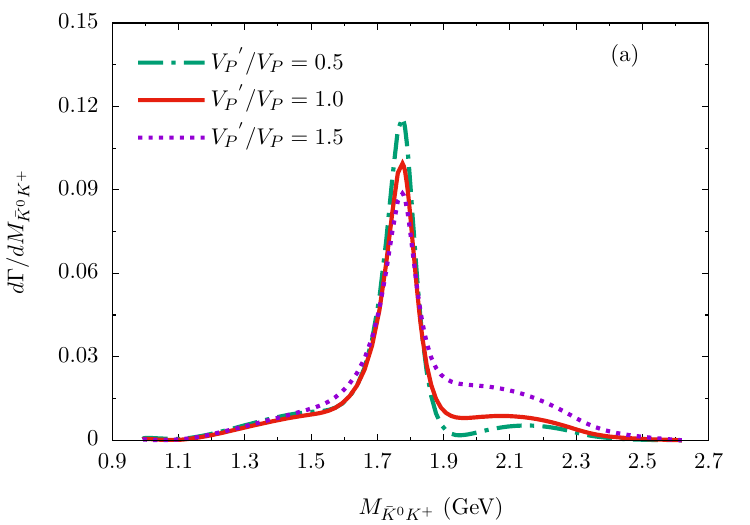}
\includegraphics[scale=0.65]{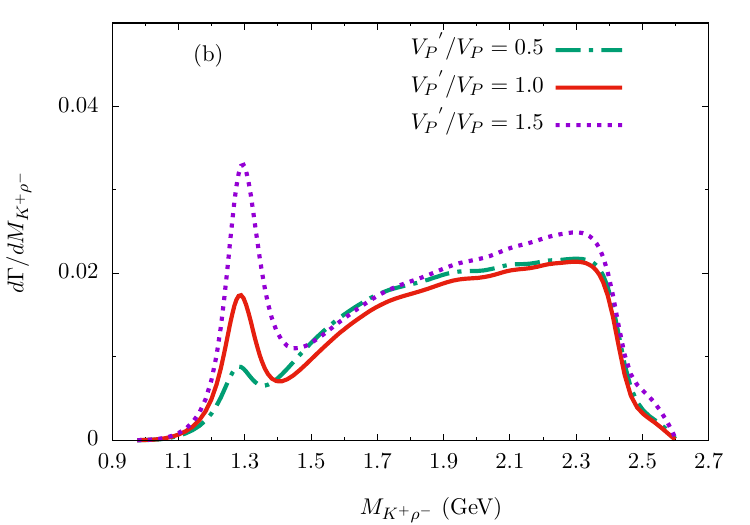}        \includegraphics[scale=0.65]{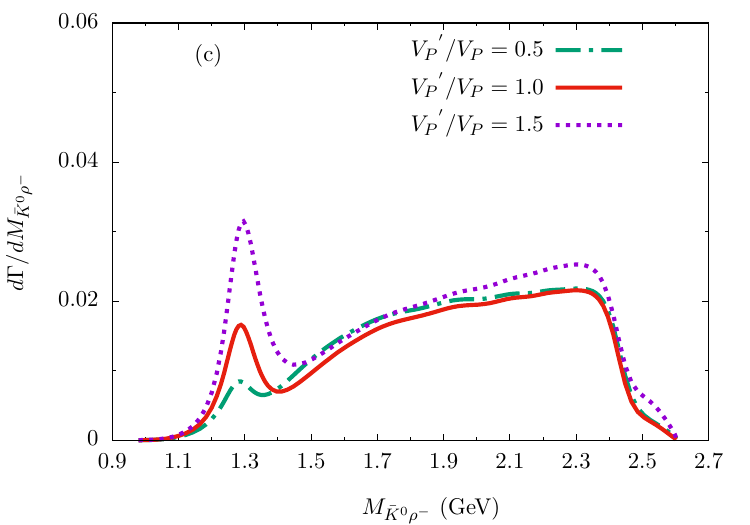}
	\caption{(a) $\bar{K}^0K^+$, (b) $K^+\rho^-$, and (c) $\bar{K}^0\rho^-$  invariant mass distributions of the process $J/\psi\rightarrow \bar{K}^0K^+\rho^-$ with $V_p^{\prime}/V_p=0.5,1.0,1.5$.}	\label{fig:vp}
\end{figure}

In Fig.~\ref{fig:kk}, we show the $\bar{K}^0K^+$ invariant mass distribution of the process $J/\psi \to \bar{K}^0K^+\rho^-$. The red-solid curve stands for the total contributions from the $a_0(1710)$ state, the axial-vector $K_1(1270)$ meson, and the $S$ wave $\bar{K}K$ interaction, while the blue-dashed curve corresponds to the contribution from the $a_0(1710)$ state. Moreover, the green-dot-dashed and purple-dotted curves correspond to the contributions from the intermediate $K_1(1270)^0$ and $K_1(1270)^-$, respectively, and orange dashed curve corresponds to the contributions from the  direction production and the $S$-wave $K\bar K$ interaction of Eq.~(\ref{eq:md}). One can find a clear peak structure around 1.8~GeV, which could be associated with the scalar $a_0(1710)$. 
However, there is no significant structure of the $a_0(980)$ near the $\bar{K}^0K^+$ threshold, because of the suppression by the phase space.
The intermediate resonances  $K_1(1270)^0$ and  $K_1(1270)^-$ give 
the smooth contributions in the region of $1.4\sim 2.4$~GeV, which is due to the fact that the $K_1(1270)$ couples to $K\rho$ in the  $S$ wave. 

However, it should be pointed out that the peak structure appearing in the $\bar{K}^0K^+$ invariant mass distribution of Fig.~\ref{fig:kk} could also manifest itself as a dip structure
 if the interference between $\mathcal{M}_a,\mathcal{M}_b,\mathcal{M}_c$, and $\mathcal{M}_d$ is different from our naive assignments explained above. For instance, if we multiply the contribution of tree diagram ``1'' of the term $\mathcal{M}_d$ of  of Eq.~(\ref{eq:fullamp}) by a phase factor $e^{i\phi}$ with $\phi=0, \pi/3, 2\pi/3, \pi$, and $4\pi/3$, we would obtain the $\bar{K}^0K^+$ invariant mass distribution shown in Fig.~\ref{fig:phi}, where one can see a dip structure around 1.8~GeV for $\phi=4\pi/3$.

Next, we have predicted the $K^+\rho^-$ and $\bar{K}^0\rho^-$  invariant mass distributions of the process $J/\psi\rightarrow \bar{K}^0K^+\rho^-$ in Figs.~\ref{fig:dw_Krho}(a) and \ref{fig:dw_Krho}(b), respectively. One can see the clear peaks of the $K_1(1270)^0$ and $K_1(1270)^-$ around 1.3~GeV. It should be stressed that the axial vector $K_1(1270)$ is predicted to have a two-pole structure, the lower pole is around 1200~MeV, coupled strongly to the $K^*\pi$ channel, and the higher pole is around 1280~MeV, coupled strongly to the $K\rho $ channels~\cite{Geng:2006yb}. 


As we discussed above, since the higher pole of $K_1(1270)$ mainly couples to the $K\rho$ channel, we have adopted the pole position and coupling constant of the higher pole in Eq.~(\ref{eq:mc}). To show the difference between the two poles of the $K_1(1270)$, we have calculated the  $\bar{K}^0K^+$, $K^+\rho^-$, and $\bar{K}^0\rho^-$ invariant mass distributions for the process  $J/\psi\rightarrow \bar{K}^0K^+\rho^-$ with the pole position and coupling constant of the lower pole of Table~\ref{tab:K1} in Eq.~(\ref{eq:mc}), as shown in Fig.~\ref{fig:lowerpole-Krho}. Because the coupling of the lower pole to $K\rho$ is smaller and the width is larger, the contribution from the lower pole of the $K_1(1270)$ is very small. For the contributions from the lower pole and the higher pole to have the same order of magnitude, we take $V'_p/{V_p}= 8.0$ in the results of Fig.~\ref{fig:lowerpole-Krho}. In the $\bar{K}^0K^+$ invariant mass distribution of Fig.~\ref{fig:lowerpole-Krho}(a), one can find the clear peak structure around 1.8~GeV. However, in the $\bar{K}^0\rho^-$ and $K^+\rho^-$ invariant mass distributions of Figs.~\ref{fig:lowerpole-Krho}(b) and \ref{fig:lowerpole-Krho}(c), only an enhancement structure appears. Thus, future measurements of $\bar{K}^0\rho^-$ and $K^+\rho^-$ invariant mass distributions could shed light on the two-pole structure of the $K_1(1270)$.

In addition, we take different ratios of  $V_p^{\prime}/V_p=0.5,1.0$, and 1.5, and show the $\bar{K}^0K^+$, $K^+\rho^-$, and $\bar{K}^0\rho^-$ invariant mass distributions for the process  $J/\psi\rightarrow \bar{K}^0K^+\rho^-$ in Figs.~\ref{fig:vp}(a)-\ref{fig:vp}(c), respectively. One can find that the peak structure of the $a_0(1710)$ in the $\bar{K}^0K^+$ invariant mass distribution remains similar for different values of $V_p^{\prime}/V_p$, and the peak structures of the $K_1(1270)$ become clearer for larger values of $V_p^{\prime}/V_p$.

\section{Summary} \label{sec:Conclusions}
Assuming the $a_0(1710)$ as a $K^*\bar{K}^*$ molecular state, we have investigated the process $J/\psi\rightarrow \bar{K}^0K^+\rho^-$ by taking into account the contribution from the $S$-wave $\omega\rho$, $K^*\bar{K}^*$, and $\phi\rho$ interactions,  as well as the contribution from the intermediate resonances $a_0(980)$ and $K_1(1270)$.  

We have predicted one peak structure around 1.8~GeV in the $\bar{K}^0K^+$ invariant mass distribution, which could be associated with the scalar meson $a_0(1710)$. However, there is no significant near-threshold enhancement structure of the $a_0(980)$ in the  $K\bar{K}$ invariant mass distribution. Furthermore, we have  also predicted the $K^+\rho^-$ and $\bar{K}^0\rho^-$ invariant mass distributions of the process $J/\psi\rightarrow \bar{K}^0K^+\rho^-$, and find clear peaks of the resonance $K_1(1270)^{0,-}$. Considering the two-pole structure of the $K_1(1270)$,  we have also calculated the results adopting parameters of the lower pole and find an enhancement structure near 1.3~GeV in the $K^+\rho^-$ and $\bar{K}^0\rho^-$ invariant mass distributions, which implies that future measurements of the $K^+\rho^-$ and $\bar{K}^0\rho^-$ invariant mass distributions could shed light on the two-pole structure of the $K_1(1270)$.

Finally, considering different weights ratio  $V_p^{\prime}/V_p=0.5,1.0$, and 1.5 of contributions from $a_0(1710)$ and $K_1(1270)$, we have shown the $\bar{K}^0K^+$, $K^+\rho^-$ and $\bar{K}^0\rho^-$ invariant mass distributions of the process $J/\psi\rightarrow \bar{K}^0K^+\rho^-$, and find that the peak structure of $a_0(1710)$ remains essentially the same. 
We hope that our theoretical predictions could be tested by the BESIII and Belle II experiments and the planned STCF in the future, and precise measurements of the process $J/\psi\rightarrow \bar{K}^0K^+\rho^-$ could shed light on the nature of the scalar $a_0(1710)$ and the axial vector $K_1(1270)$.


\section*{Acknowledgments}
L.S.G and J.J.X acknowledge support from the National Key R\&D Program of China under Grant No. 2023YFA1606700. 
This work is supported by the Natural Science Foundation of Henan under Grant No. 222300420554 and No. 232300421140.
This work is partly supported by the National Natural
Science Foundation of China under Grants Nos. 12075288, 11975041,  11961141004, 12361141819, and 12192263, the Project of Youth Backbone Teachers of Colleges and Universities of Henan Province (2020GGJS017),  the Open Project of Guangxi Key Laboratory of Nuclear Physics and Nuclear Technology, No. NLK2021-08, and the Central Government Guidance Funds for Local Scientific and Technological Development, China (No. Guike ZY22096024). It is also partly supported by the Youth Innovation Promotion Association CAS.

\end{document}